# Current driven "plasmonic boom" instability in three-dimensional gated periodic ballistic nanostructures

G. R. Aizin[1*], J. Mikalopas[1], and M. Shur[2†]

[1] *Kingsborough College, The City University of New York, Brooklyn, New York 11235, USA*
[2] *Rensselaer Polytechnic Institute, Troy, New York 12180, USA*

A new approach of using distributed transmission line analogy for solving transport equations for ballistic nanostructures is applied for solving the three dimensional problem of the electron transport in gated ballistic nanostructures with periodically changing width. The structures with the varying width allow for modulation of the electron drift velocity while keeping the plasma velocity constant. We predict that in such structures biased by a constant current, a periodic modulation of the electron drift velocity due the varying width results in the instability of the plasma waves if the electron drift velocity to plasma wave velocity ratio changes from below to above unity. The physics of such instability is similar to that of the sonic boom, but, in the periodically modulated structures, this analog of the sonic boom is repeated many times leading to a larger increment of the instability. The constant plasma velocity in the sections of different width leads to the resonant excitation of the unstable plasma modes with the varying bias current. This effect (that we refer to as the super plasmonic boom condition) results in a strong enhancement of the instability. The predicted instability involves the oscillating dipole charge carried by the plasma waves. The plasmons can be efficiently coupled to the terahertz (THz) electromagnetic radiation due to the periodic geometry of the gated structure. Our estimates show that the analyzed instability should enable powerful tunable terahertz electronic sources.

*gaizin@kbcc.cuny.edu
†shurm@rpi.edu

# I. INTRODUCTION

The plasma wave propagation in the two-dimensional electron gas is strongly affected by the electron drift. At the values of the drift velocity smaller that the plasma velocity, the Doppler effect leads to the plasma wave instability [1,2]. When the drift velocity reaches the plasma velocity, the electron flow is "choked" leading to the current saturation [3]. The transition from the sub-plasmonic drift velocity to the super-plasmonic drift velocity should be accompanied by the "plasmonic boom" similar to the sonic boom. This analogy is due to the hydrodynamic equations describing the plasma wave of the small amplitude being identical to those describing the sound waves. The plasmonic boom effect can be used for exciting plasmons with rapidly increasing amplitude in the periodically modulated two-dimensional electron gas (2DEG) [4,5]. Since the plasma frequency in the periodically modulated 2DEG structures is typically in the THz range, this instability should lead to the emission of the THz radiation enabling a new type of the THz electronic sources. As shown in this paper, the instability is resonantly enhanced if the plasma velocity is the same in all device regions (we refer to this condition as a super plasmonic boom.)

Developing an efficient electronic THz source is one of the key challenges to be met for closing the famous THz gap [6]. The existing electronic sources use Gunn diodes with frequency multiplication by Schottky diodes [7] and InGaAs based High Electron Mobility Transistor Integrated Circuits [8]. These and other similar electronic sources suffer from low power, low efficiency and high cost. Using the plasma wave instabilities in ballistic Field Effect Transistors (FETs) proposed in [1,9] has a promise of developing more efficient THz sources. However, the observed THz radiation [10-12] has mostly been broadband until recently, when the proposed arrays of the ballistic FETs [13,14] have been implemented and improved to include Asymmetric Digital Grated Gate structures [15]. Nevertheless, the goal of reaching 1 mW at 1 THz using plasmonic sources has not been reached yet.

A recent proposal was to use a grating gate periodic structure with two sections in each period, such that the electron velocity has the value between the values of the plasma wave velocity in these sections [4,5]. In such a structure, the plasma waves behave similar to the sound waves emitted during the sonic boom, when a jet crosses the sound barrier, except that such transition occurs many times over. In Refs [4,5], the multi-gated structure with two sections having different electron densities was proposed to modulate the plasma velocity.

In this paper, we develop a theory of the 'plasmonic boom" instability in a periodically modulated 2D electron channel. We propose and analyze a more general structure, where either the periodic modulation of the electron velocity or the plasma frequency or both achieve the repeated "plasmonic boom" conditions. Our approach allows us to analyze the new structures with a periodic modulation of the device width. In these structures biased by a constant current the electron drift velocity periodically changes but the plasma velocity remains constant. The constant plasma velocity in the sections of the different width leads to the resonant excitation of the unstable plasma modes when the bias current is tuned thus strongly enhancing the instability (the super plasmonic boom). Our estimates show that the analyzed instability should enable powerful tunable terahertz electronic sources. The proposed structure might be the most practical implementation of a periodic THz source, see Figure 1.

The structure shown in Figure 1 consists of the alternating 2D strips with two different widths $W_1 < W_2$ and lengths $L_1$ and $L_2$. The metal gate positioned above the 2D channel control the electron density in the strips and allows tuning of the plasma wave velocity. We also assume that a dc current flows between the source and the drain. This structure represents a periodic plasmonic medium forming a 1D plasmonic crystal [4,5,16-22]. In this paper, we demonstrate that constant electron drift qualitatively changes the plasmonic crystal spectrum and may result in the instability of the drifting plasma modes different from the plasma instability regimes described in [1] and [23].

## II. BASIC EQUATIONS

We will describe plasma oscillations in the 2D electron gas in the presence of a dc electric current within the hydrodynamic model. In this model, the local electron density $n(x,t)$ and velocity $v(x,t)$ obey the Euler and continuity equations

$$\frac{\partial v}{\partial t} + v\frac{\partial v}{\partial x} = \frac{e}{m^*}\frac{\partial \varphi}{\partial x}$$

$$\frac{\partial n}{\partial t} + \frac{\partial (nv)}{\partial x} = 0,$$

(1)

where we assumed that the plasma wave in the 2D layer ($z = 0$) propagates in the $x$-direction between the source and the drain. Here $\varphi(x, z = 0, t)$ is the electric potential in the 2D plane, $-e$ and $m^*$ are the electron charge and effective mass, respectively. We neglect collisional damping in the Euler equation assuming the ballistic transport. Eqs. (1) could be linearized for small fluctuations of the electron density and velocity: $n = n_0 + \delta n$, $v = v_0 + \delta v$, where $n_0$ is the equilibrium electron density in the 2D channel and $v_0$ is the electron drift velocity due to dc source-drain electric current. We also assume that the system fluctuations of the electron density $\delta n$ and electric potential $\delta \varphi$ in the gated 2D electron channel are connected as $-e\delta n = C\delta\varphi$, where $C = \varepsilon/4\pi d$ is the capacitance per unit area between the 2D channel and the metal gate separated by the distance $d$, $\varepsilon$ is the dielectric constant of the barrier layer between the metal and the 2D channel. This assumption is justified if $d$ is much less than the plasmon wavelength. The solution of Eqs. (1) (linearized with respect to $\delta n, \delta v \propto exp(-iqx + i\omega t)$) is

$$I_\omega = I_1 e^{-iq_1 x} + I_2 e^{-iq_2 x}$$

$$V_\omega = \frac{1}{CW}\left(\frac{I_1}{v_0+v_p}e^{-iq_1 x} + \frac{I_2}{v_0-v_p}e^{-iq_2 x}\right),$$

(2)

where $I_\omega = W\delta j_\omega = -eW(v_0\delta n_\omega + n_0\delta v_\omega)$ is the total current in the 2D channel of width $W$ and $V_\omega \equiv \delta\varphi_\omega(x)$ is the voltage distribution in the plasma wave (both taken at frequency $\omega$). The plasmonic wave vectors $q_{1,2}$ are determined as $q_{1,2} = \omega/(v_0 \pm v_p)$ where $v_p = \sqrt{e^2 n_0/m^* C}$ is velocity of the gated acoustic 2D plasmon in the absence of the drift [24]. Drifting plasmon in the gated channel also has linear dispersion but with Doppler shifted wave velocity due to constant electron drift [1]. Constants $I_{1,2}$ in Eqs. (2) are determined by the boundary conditions.

The total power carried by the drifting plasma wave includes the electromagnetic power and the kinetic power due to the drift of the electrons oscillating in the wave. The average complex power $P_\omega(x)$ carried by the drifting plasmon in the $x$-direction can be written as

$$P_\omega(x) = W \int_{-\infty}^{\infty} S_x dz - \frac{m^* v_0}{2e} \delta v_\omega I_\omega^* , \qquad (3)$$

where $S_x = -(c/8\pi) E_{z,\omega} B_{y,\omega}^*$ is the $x$-component of the complex Poynting vector averaged over the THz period. The first term in Eq. (3) describes the electromagnetic power. The second one represents the kinetic power and vanishes at $v_0 = 0$. The electric ($E_{z,\omega}$) and magnetic ($B_{y,\omega}$) fields in the plasma wave as well as the kinetic power can be expressed in terms of the voltage $V_\omega$ and the current $I_\omega$ from Eqs. (2). After the integration Eq. (3) yields

$$P_\omega = \frac{1}{2} \left( \frac{v_p^2 - v_0^2}{v_p^2} V_\omega + \frac{v_0}{CW v_p^2} I_\omega \right) I_\omega^* . \qquad (4)$$

At $v_0 = 0$ the standard expression $P = VI^*/2$ for the power flow in the plasmonic waveguide in the limit of the strong gate screening is recovered. The electron drift effectively modifies the voltage distribution in the plasmonic waveguide by adding the so-called kinetic voltage: $-m^* v_0 \delta v$, first introduced for description of the electron beam waves in tubes [25]. Expression for the power flow in Eq. (4) reduces to its standard form after defining an effective voltage

$$V_\omega^{eff} = \frac{v_p^2 - v_0^2}{v_p^2} V_\omega + \frac{v_0}{CW v_p^2} I_\omega . \qquad (5)$$

According to Eqs. (2) and (5), the values of $V_\omega^{eff}$ and $I_\omega$ at the opposite boundaries of the 2D electron strip of length $\ell$ ($x = 0, \ell$) are connected via the transfer matrix $\hat{t}$:

$$\begin{pmatrix} V_\omega^{eff}(0) \\ I_\omega(0) \end{pmatrix} = \hat{t} \begin{pmatrix} V_\omega^{eff}(\ell) \\ I_\omega(\ell) \end{pmatrix}, \qquad (6)$$

where

$$\hat{t} = e^{-i\omega \frac{v_0}{v_p^2} \Theta} \begin{pmatrix} \cos\Theta & \frac{i}{WCv_p} \sin\Theta \\ iWCv_p \sin\Theta & \cos\Theta \end{pmatrix}, \quad \Theta = \frac{\omega \ell v_p}{v_p^2 - v_0^2} . \qquad (7)$$

The dispersion relation for the drifting 1D plasmonic crystal formed in the structure of Figure 1 depends on the boundary conditions between the strips 1 and 2 in the crystal elementary cell. In the limit of a strong gate screening, the 2D channel in each strip can be considered as a plasmonic transmission line (TL) supporting 2D TEM plasma waves [20,21]. In the absence of drift, the continuity of the current and voltage at the boundary between the strips represents the standard TL boundary conditions providing the continuity of the power flow through the boundary. A finite drift breaks the reciprocity of the TL due to the different wave velocities of the plasmons propagating in the opposite directions. To preserve the continuity of the power flow, we assume the continuity of the current $I_\omega$ together with the continuity of the effective voltage $V_\omega^{eff}$ defined in Eq. (5). For these boundary conditions, the values of $I_\omega$ and $V_\omega^{eff}$ at the opposite sides of the crystal elementary cell are connected by the transfer matrix $\hat{t}_2 \hat{t}_1$ where $\hat{t}_i$, $i = 1,2$ are the transfer matrices defined in Eq. (7) for the strips 1 and 2. In the translationally invariant periodic plasmonic medium, the dispersion equation for the 1D drifting plasmonic crystal can be found using the Bloch theorem and solving the resulting 1D Kronig-Penney problem [26]

$$\cos(kL + \omega T) = \cos \omega T_1 \cos \omega T_2 - \frac{1}{2}\left(\gamma + \frac{1}{\gamma}\right) \sin \omega T_1 \sin \omega T_2 , \qquad (8)$$

Here $\gamma = \frac{W_1 v_{p1} d_2}{W_2 v_{p2} d_1}$, $L = L_1 + L_2$ is the crystal lattice constant, $k \in [-\pi/L, \pi/L]$ is the plasmon Bloch wave vector, and indices 1 and 2 refer to the strips 1 and 2, respectively. Parameters $T$ and $T_i$, $i = 1,2$ are defined as [5]

$$T_i = \frac{L_i v_{pi}}{v_{pi}^2 - v_{0i}^2}, \quad i = 1,2 \ ; \quad T = \sum_{i=1,2} \frac{v_{0i}}{v_{pi}} T_i \qquad (9)$$

Eq. (8) generalizes the dispersion equation for the 1D drifting plasmonic crystal found in Ref. [5] to the case of periodically changing strip width $W$ and the gate-to-channel distance $d$. Parameter $\gamma \leq 1$ in Eq. (8) describes the modulation depth of the plasmonic medium.

### III. RESULTS AND DISCUSSION

We will now consider Eq. (8) in the limit of strong modulation, $\gamma \ll 1$, and look for solution in the form of the power asymptotic series $\omega = \sum_{p=0}^\infty \omega_p \gamma^p$. Substituting this expansion into Eq. (8) and combining terms of the same order, we find for the first two terms of the asymptotic series

$$\omega_m^{(i)} = \omega_{0,m}^{(i)} + \frac{2\left[(-1)^{m+1} \cos\left(kL + \omega_{0,m}^{(i)} T\right) + \cos \omega_{0,m}^{(i)} T_j\right]}{T_i \sin \omega_{0,m}^{(i)} T_j} \gamma + O(\gamma^2);$$
$$\omega_{0,m}^{(i)} = \frac{\pi m}{|T_i|}, \qquad i,j = 1,2 \ i \neq j, \ m = 1,2,\ldots \qquad (10)$$

Frequencies $\omega_{0,m}^{(i)}$ in Eq. (10) are the frequencies of the drifting plasmons confined in the cavity of length $L_i$ with the symmetric boundary conditions. In Figure 2, we plot the frequencies of the first three quantized plasmonic levels in the strips 1 and 2 as a function of the electron drift velocity in the narrow strip 1, $v_{01}$. In this Figure, we assume that both strips 1 and 2 have the same parameters except for the width so that $\gamma = W_1/W_2 < 1$ and the electron drift velocity in the wide strip 2: $v_{02} = \gamma v_{01}$.

It follows from Eq. (10) that the quantized plasmonic energy levels in the identical strips (1 or 2) are weakly coupled and broadened into the narrow plasmonic bands at $\gamma \ll 1$. In this limit, strips 1 and 2 form two independent plasmonic sublattices. The plasmon frequencies in the bands in Eq. (10) are real. Hence, no instability occurs at any value of the electron drift velocity.

Points where $\omega_{0,m}^{(1)} = \omega_{0,p}^{(2)}$, $m, p = 0,1,2, \ldots$ present special interest. In the absence of a dc drift, the band gaps in the plasmonic crystal spectrum vanish in these points, and plasma wave propagates through the entire crystal in a resonant manner [21]. The dispersion law for the drifting plasmon in these transparency points cannot be found from Eq. (10) because all terms in the asymptotic series used in this equation diverge, and an alternative asymptotic expansion is developed below.

Let $\omega_{0,m}^{(1)} = \omega_{0,p}^{(2)} \equiv \omega_{0,mp}$. At $\gamma \ll 1$ solution of Eq. (8) has the form $\omega_{mp} = \omega_{0,mp} + \Delta\omega$ where $\Delta\omega \to 0$ at $\gamma \to 0$. We expand Eq. (8) into quadratic polynomial with respect to $\Delta\omega$ and look for $\Delta\omega$ in the form of an asymptotic series $\Delta\omega = \sum_{k=0}^{\infty} \alpha_k \gamma^{\delta_k}$, $\delta_{k+1} > \delta_k$. Coefficients $\alpha_k$ and $\delta_k$ can be found by the standard Newton diagram method used for finding asymptotic expansion of the polynomial roots [27]. Our calculations yield

$$\omega_{mp}^{(\pm)} = \omega_{0,mp} \pm \frac{2\gamma^{1/2}}{\sqrt{T_1 T_2}} \begin{cases} \sin\left(\frac{kL+\omega_{0,mp}T}{2}\right), m+p \text{ even} \\ \cos\left(\frac{kL+\omega_{0,mp}T}{2}\right), m+p \text{ odd} \end{cases} + \frac{(-1)^{m+p} T \sin(kL+\omega_{0,mp}T)}{T_1 T_2} \gamma + O(\gamma^{3/2}) \quad (11)$$

It follows from Eq. (11) that at $T_1 T_2 > 0$, the resonant coupling of the quantized plasmonic levels in the adjacent non-identical strips 1 and 2 splits each unperturbed degenerate plasmonic level into the two plasmonic bands described by the second term in Eq. (11). If $T_1 T_2 < 0$, the degenerate plasmonic level broadens into one narrow plasmonic band described by the third term in Eq. (11). The second term in this equation becomes purely imaginary and corresponds to either unstable or decaying plasmon modes. The instability increment depends on the Bloch wave vector. From Eq. (9) it follows that the instability occurs when $v_{01} > v_{p1}$ ($v_{01} < v_{p1}$) but $v_{02} < v_{p2}$ ($v_{02} > v_{p2}$). These inequalities constitute the necessary conditions for the repeated "sonic boom" with the dc current flowing in the structure. In Figure 2, the stable and unstable transparency points are marked by the open and closed circles, respectively. One can also show that at $\omega_{0,m}^{(1)} \neq \omega_{0,p}^{(2)}$ but $\left|\omega_{0,m}^{(1)} - \omega_{0,p}^{(2)}\right| \to 0$ the interaction between nearly degenerate plasmonic levels in the strips 1 and 2 results in either two split plasmonic bands with real eigenvalues or one band with unstable and decaying branches dependent on the value of $k$ in the Brillouin zone.

These analytical results are confirmed by the direct numerical solution of Eq. (8). For the structure shown in Fig. 1, we choose $v_{p1} = v_{p2} \equiv v_p$ and $d_1 = d_2$ so that $\gamma = W_1/W_2 < 1$ and $v_{02} = \gamma v_{01}$. We also assume that $L_1 = L_2 = L/2$ and use dimensionless units for the complex plasma frequency $(\omega' + i\omega'')/\omega_0$ where $\omega_0 = 2v_p/L$. The dimensionless drift velocity in the narrow strips is defined as $\tilde{v}_0 = v_{01}/v_p$.

Figs. 3-6 show the results of the numerical solution of Eq. (8) for $\gamma = 0.1$. Fig. 3 shows the drifting plasmonic crystal spectrum for $\tilde{v}_0 = 0.43$. At this value of the drift velocity, $\omega_{0,m}^{(1)} \neq \omega_{0,p}^{(2)}$ at any $m, p \gtrsim 1$, and the low energy spectrum consists of the two sets of the plasmonic bands $\omega_m^{(1)}$ and $\omega_m^{(2)}$, $m = 1,2, ...$ formed due to the resonant coupling of the plasmon energy levels $\omega_{0,m}^{(1)}$ and $\omega_{0,m}^{(2)}$ in the strips 1 and 2, respectively, as described by Eq. (10). One additional low energy solution appears in the transparency point $\omega_{0,0}^{(1)} = \omega_{0,0}^{(2)}$. This solution can be interpreted as a lattice acoustic plasmon similar to the acoustic phonons in the atomic crystal lattice and corresponds to the $\omega_{00}^{(\pm)}$ modes in Eq. (11). Since at given value of the drift velocity $T_1 T_2 > 0$, there is no imaginary part in the frequencies $\omega_{00}^{(\pm)}$.

At $\gamma = 0.1$, the instability occurs in the range of $1 < \tilde{v}_0 < 10$. Fig. 4 shows the numerically found plasmonic spectrum for $\tilde{v}_0 = 1.12$ when $\omega_{0,4p}^{(1)} = \omega_{0,p}^{(2)}$, $p = 0,1,2, ...$. The real part of the plasma frequencies, $\omega'/\omega_0$, is plotted in Fig. 4a. The plasmonic spectrum consists of the stable bands with purely real frequencies formed by the coupled strips 1, $\omega_m^{(1)}$, and unstable bands, $\omega_{mp}^{(\pm)}$, in the transparency points where $\omega_{0,m}^{(1)} = \omega_{0,p}^{(2)}$. Fig. 4a shows the two of these unstable bands. The instability increment, $|\omega''|/\omega_0$, in these bands depends on the plasmonic Bloch wave vector as shown in Fig. 4b. These results correlate very well with asymptotic analytical formulas in Eqs. (10) and (11).

All the quantized plasmonic levels in strips 1 and 2 are perfectly matched if $|T_1| = |T_2|$, see Eqs. (9) and (10). This last equation represents the super resonance condition when the plasmonic crystal becomes unstable at any plasma frequency $\omega_{mp}^{(\pm)}$, provided that $T_1 T_2 < 0$ (the super plasmonic boom). For the structure considered here, it happens at $\tilde{v}_0 = \sqrt{2/(1+\gamma^2)} \approx 1.41$. In Fig. 2, the super resonance condition is marked by the dashed vertical lines with red arrow. Plasma frequencies and instability increments for this totally unstable plasmonic crystal are shown in Figs. 5a and 5b, respectively.

The results presented in Figs. 4 and 5 indicate that the instability has a resonant character and occurs every time when there is a perfect plasmonic level matching between different strips in the transparency points. In this case, the plasma modes are unstable at any value of the plasmonic Bloch wave vector $k$. However if the level mismatch is small the instability does not completely vanish but occurs at some intervals of $k$ in the Brillouin zone. In Fig. 6, we plotted the plasmon dispersion curves for $\tilde{v}_0 = 1.71$ when $\omega_{0,p}^{(1)} = \omega_{0,2p}^{(2)}$, $p = 0,1,2, ...$. Two unstable modes $\omega_{12}^{(\pm)}$ and $\omega_{00}^{(\pm)}$ are shown in Fig. 6a with the corresponding instability increments shown in Fig. 6b. The unstable mode $\omega_{12}^{(\pm)}$ is the result of the resonant coupling of the plasmonic levels $\omega_{0,1}^{(1)}$ and $\omega_{0,2}^{(2)}$. When the drift velocity $\tilde{v}_0$ changes, these two levels shift differently and decouple. The inset shows the dispersion curves for this mode at $\tilde{v}_0 = 1.65$. The instability disappears at some interval of $k$, where instead of one unstable band two split stable bands emerge. When level mismatch increases the region of stability expands and finally the unstable band $\omega_{12}^{(\pm)}$ transforms into two stable bands $\omega_1^{(1)}$ and $\omega_2^{(2)}$ described by Eq. (10).

One should also point out that the low frequency acoustic mode $\omega_{00}^{(\pm)}$ in the plasmonic crystal lattice remains unstable at any value of $\tilde{v}_0$ within the instability window $1 < \tilde{v}_0 < 1/\gamma$. This result follows from the asymptotic expansion in Eq. (11) and is confirmed by the numerical simulations shown in Figs. 3-6.

As seen from Figs. 4-6, the predicted instability increment is of the order of $1/\omega_0$, where $\omega_0$ is the plasma frequency. This corresponds to the gain coefficient $g \sim \omega_0/\tilde{v}_p$, where $\tilde{v}_p = \partial\omega/\partial k$ is the plasma group velocity in the plasmonic crystal energy band, $\tilde{v}_p \sim \gamma v_p$. For the plasma frequency of 3 THz ($\omega_0 \sim 2\mathrm{x}10^{13}$ s$^{-1}$) and $\tilde{v}_p \sim 10^5$ m/s, the gain $g \sim 2\mathrm{x}10^8$ m$^{-1}$. Demanding $g \gg 1$ for the efficient generation, we need structures with the length on the order of one micron. For a ballistic structure, the contact resistance on the order of 0.5 Ωmm limits the device resistance. Assuming a typical radiation resistance of 300 Ω and the device width of 10 μm we estimate the current carrying capability on the order of $I_{max}$ ~ 30 mA for the electron velocity ~$2\times10^5$ m/s and the 2D electron density of $10^{13}$ cm$^{-2}$. Assuming the current swing of 0.5 $I_{max}$, we obtain the power of 80 mW with efficiency of approximately 20%.

## IV. CONCLUSIONS

The results of the analytical theory and numerical simulations show that the plasma waves in a 1D plasmonic crystal become unstable when the electron drift velocity changes from the value smaller than the plasma velocity to the value larger than plasma velocity. This can be achieved by changing either the drift velocity or the plasma wave velocity in the strips constituting the plasmonic crystal. The qualitative physics of the instability is similar to the physics of the sonic boom, which occurs when a jet airliner crosses the sound barrier. The difference is that such "plasmonic boom" is repeated many times in the periodic plasmonic structure leading to a much stronger instability. Further enhancement of the instability occurs due to the resonant excitation of the unstable plasma modes including the super resonance condition when all plasma modes become unstable (the super plasmonic boom). Another advantage of this approach is that the plasmonic crystal can efficiently couple with THz electromagnetic radiation. In our analysis, we neglected the electron collisions with impurities and lattice vibrations. The modern silicon VLSI fabrication techniques reached a feature size of 10 nm in 2015 [28], which is smaller than the mean free path in Si at room temperature (~ 30nm). This makes the ballistic plasmonic crystal analysis to be realistic as the first approximation at room temperature and to be a very good approximation at cryogenic temperatures. High values of the electron mobility in graphene (up to 200,000 cm$^2$/V-s at room temperature [29]) make this material a good candidate for the THz plasmonic crystal application [30]. Therefore, we believe that the THz generation mechanism proposed in this paper should enable a new generation of efficient and compact THz sources.

## ACKNOWLEDGMENTS

The work at RPI was supported by the US Army Research Laboratory under ARL MSME Alliance (Project manager Dr. Meredith Reed).

## Figure captions

Figure 1. Schematic diagram of the 2D transistor structure with modulated width.

Figure 2. First three quantized plasmonic levels in the non-interacting plasma cavities formed in strips 1 ($\omega_{0,p}^{(1)}$, red lines) and strips 2 ($\omega_{0,m}^{(2)}$, blue lines) as a function of the electron drift velocity in strips 1. Open (solid) circles indicate stable (unstable) transparency points. Vertical dashed lines marked with red arrow correspond to the super resonance condition.

Figure 3. Energy band spectrum of the drifting plasmonic crystal when the electron drift velocity in both strips 1 and 2 is less than the plasma wave velocity: $v_{01}, v_{02} < v_p$. Here $v_{01} = 0.43 v_p, v_{02} = 0.1 v_{01}$. Two lowest stable plasmonic bands formed due to the resonant coupling of the quantized plasmonic levels in strips 1 ($\omega_m^{(1)}$) and strips 2 ($\omega_m^{(2)}$) are shown. Stable bands $\omega_{00}^{(\pm)}$ correspond to the lattice acoustic plasmon as described in the text.

Figure 4. Energy band spectrum of the drifting plasmonic crystal when the electron drift velocity is within the instability range: $v_{02} < v_p < v_{01}$. Here $v_{01} = 1.12 v_p, v_{02} = 0.1 v_{01}$. (a) Plasmonic band frequencies in the stable bands ($\omega_m^{(1)}$, solid black lines) and unstable bands ($\omega_{mp}^{(\pm)}$, red circles and blue squares); (b) Instability increments in the unstable plasmonic bands

Figure 5. Totally unstable energy band spectrum of the drifting plasmonic crystal at $v_{01} = 1.41 v_p, v_{02} = 0.1 v_{01}$ corresponding to the resonant coupling of all quantized plasmonic levels in strips 1 and 2. The first four unstable bands are shown.
(a) Plasmonic band frequencies; (b) Instability increments.

Figure 6. Energy band spectrum of the drifting plasmonic crystal at the electron drift velocity within the instability range: $v_{01} = 1.71 v_p, v_{02} = 0.1 v_{01}$. (a) Plasmonic band frequencies in the stable bands (solid black lines) and unstable bands (red circles and blue squares); (b) Instability increments in the unstable bands. Inset: the $\omega_{12}^{(\pm)}$ unstable band at $v_{01} = 1.65 v_p$ showing the evolution of the unstable band with the changing drift velocity as described in the text.

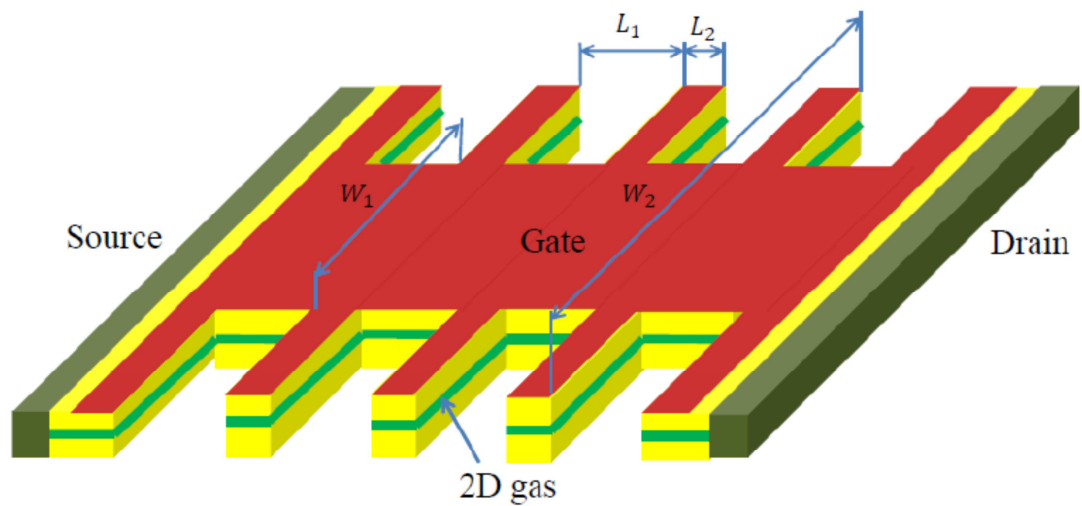

Figure 1

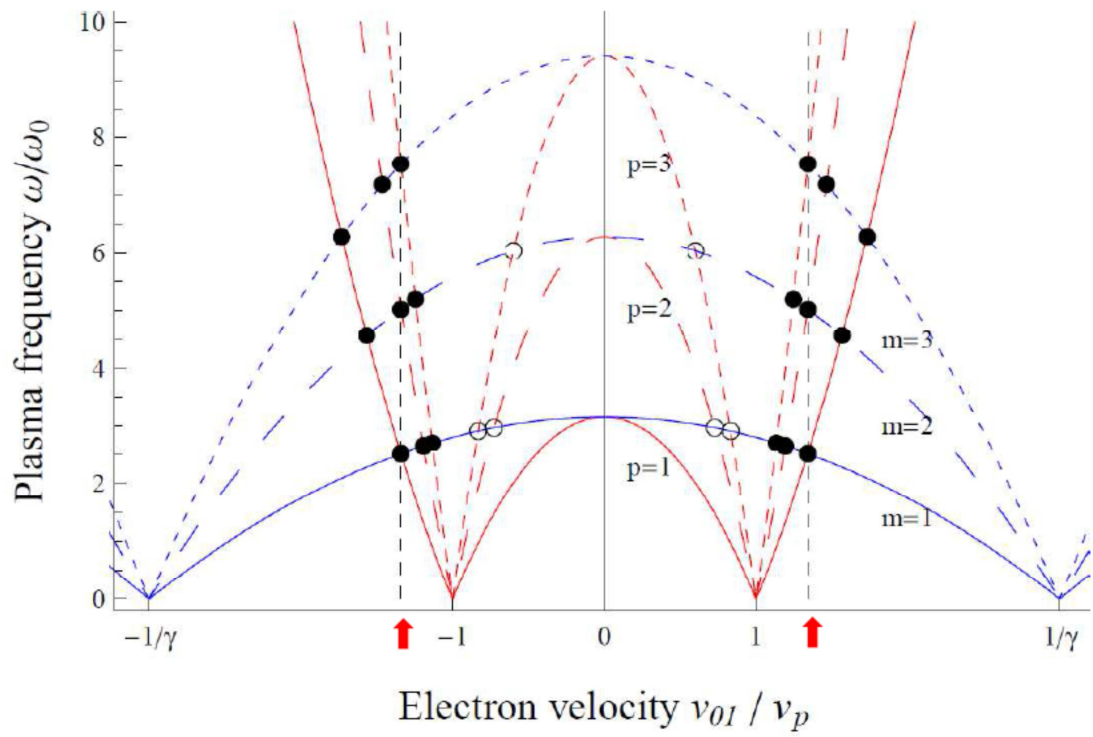

Figure 2

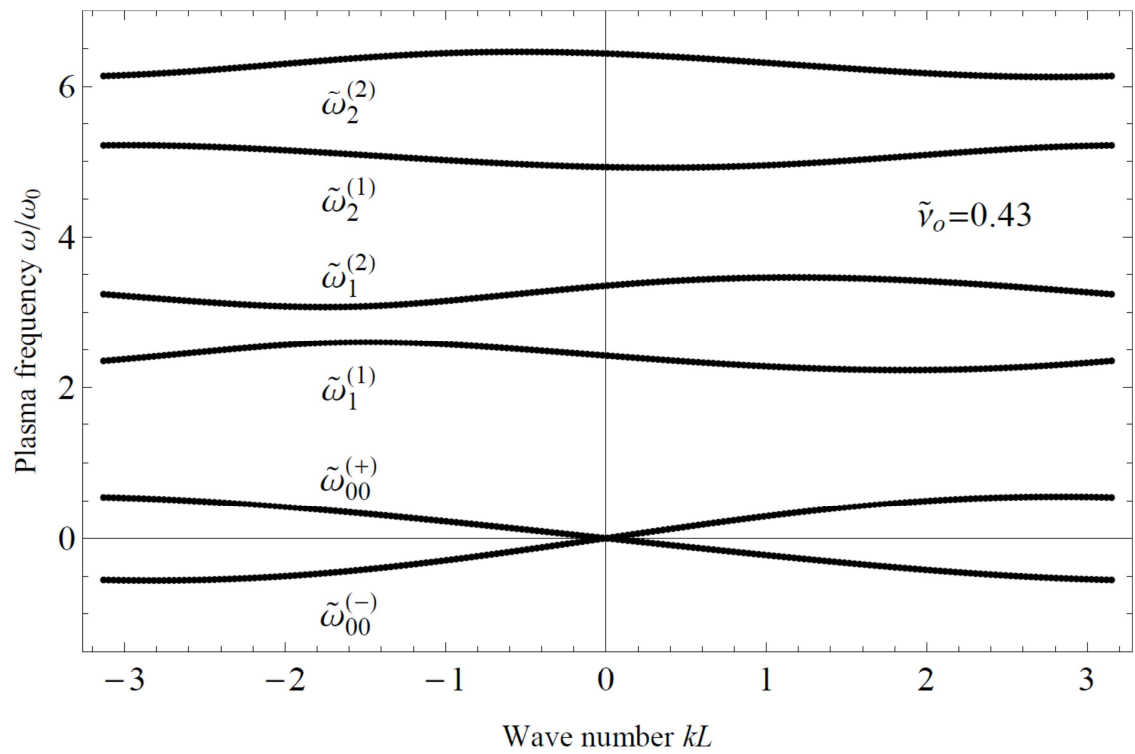

Figure 3

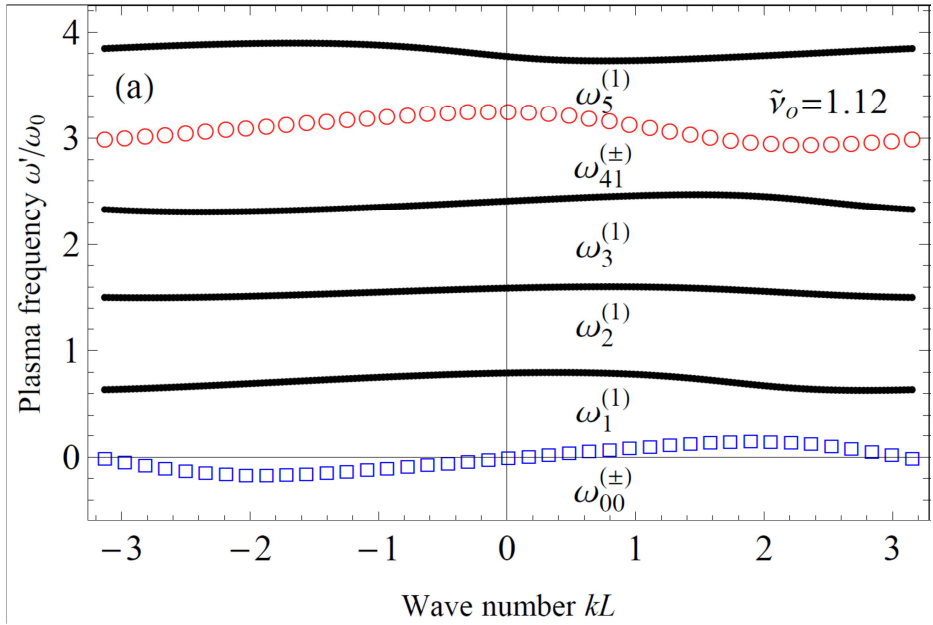

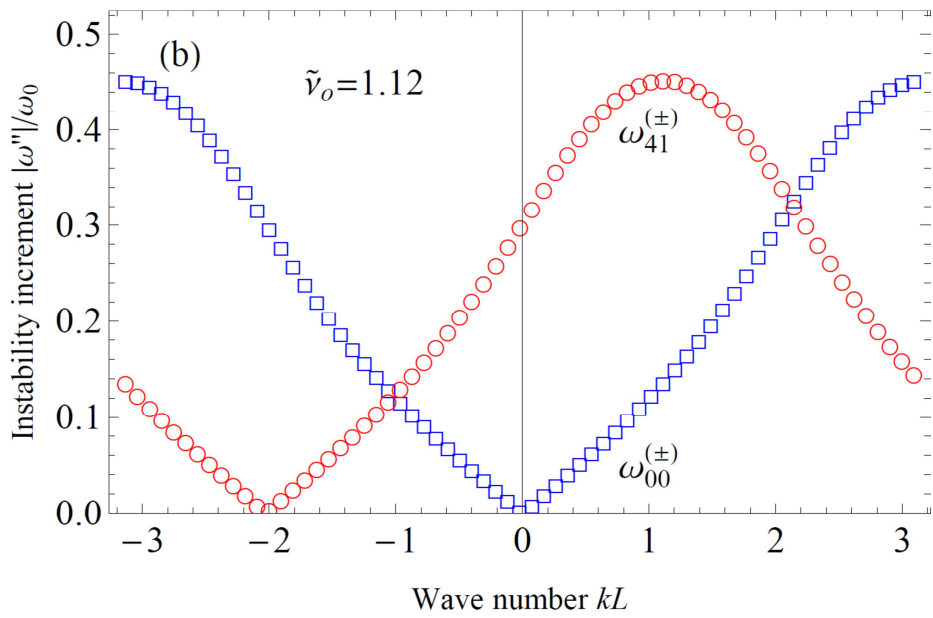

Figure 4

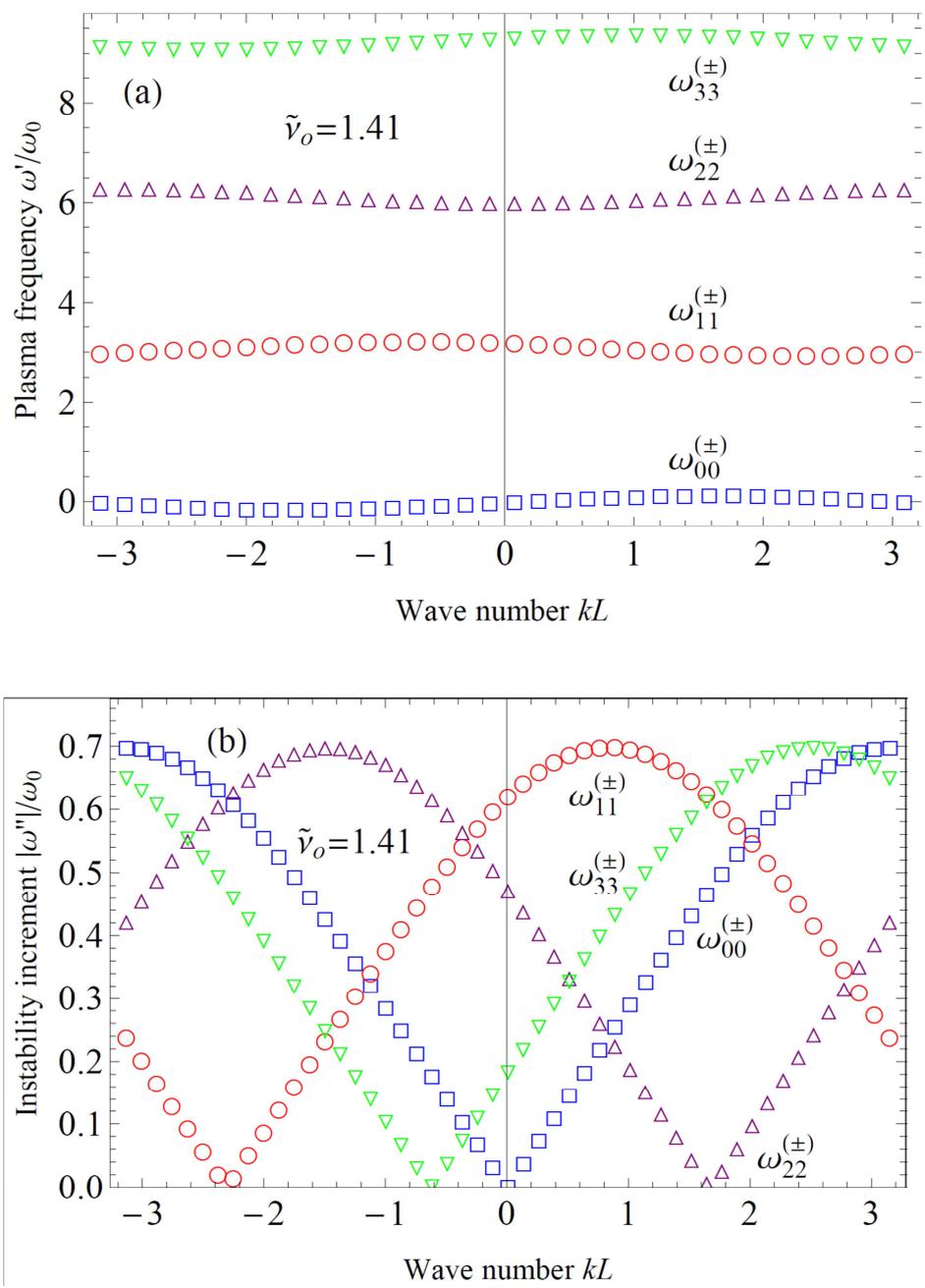

Figure 5

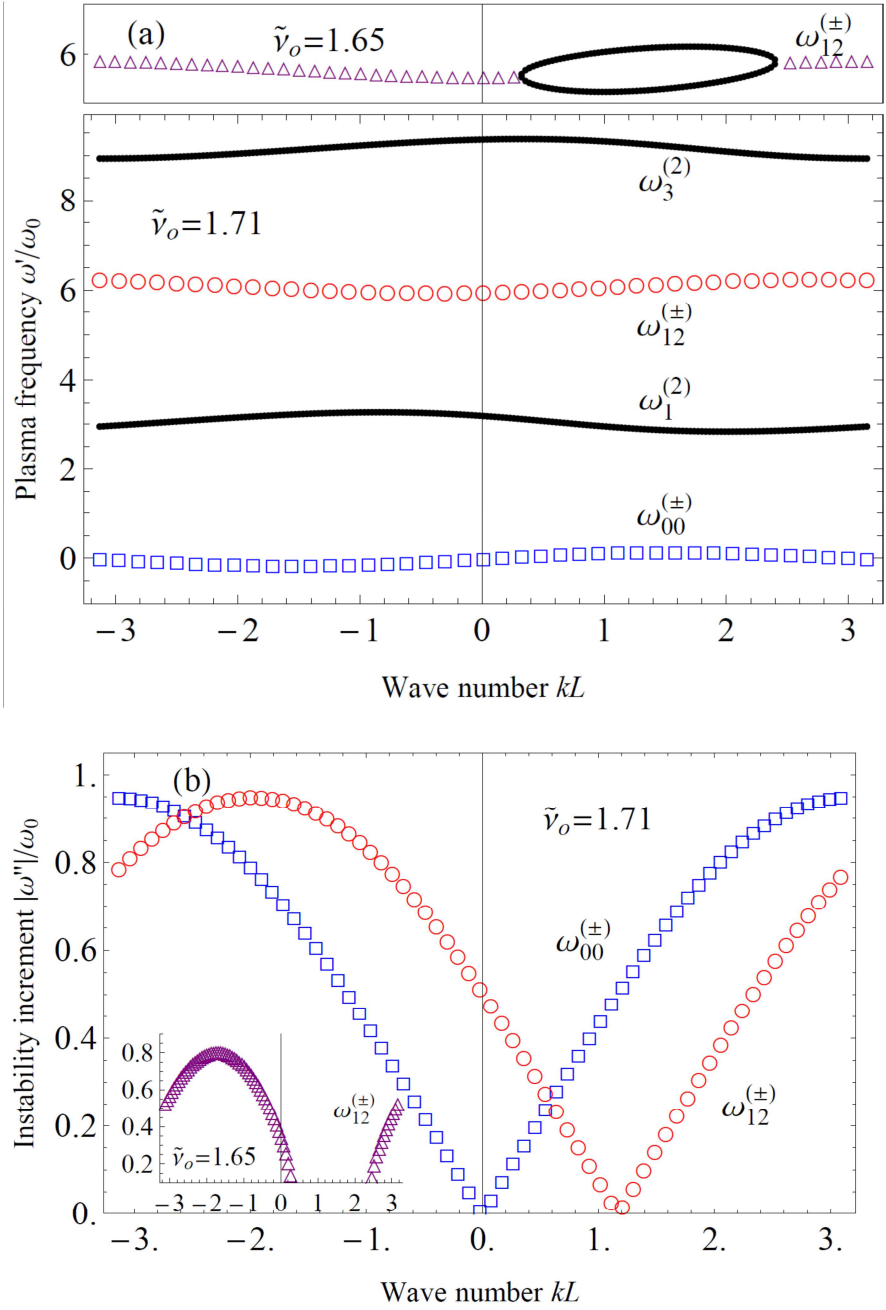

Figure 6